\documentclass[sigconf]{acmart}
\usepackage{graphicx}

\usepackage{booktabs} 
\usepackage{subfigure}
\usepackage{url}
\usepackage{color}

\copyrightyear{2018}
\acmYear{2018}
\setcopyright{none}
\acmConference{SFMA'18}{April 23, 2018}{Porto, Portugal}

\newcommand{\ctext}[1]{\raise0.2ex\hbox{\textcircled{\scriptsize{#1}}}}

\begin{document}
\title{Reactive NaN Repair for Applying Approximate Memory to Numerical Applications}
\author{Shinsuke Hamada}
\affiliation{%
  \institution{Tokyo University of Agriculture\\ and Technology}
}
\email{hamada@namikilab.tuat.ac.jp}

\author{Soramichi Akiyama}
\affiliation{%
  \institution{National Institute of Advanced Industrial Science and Technology}
}
\email{s.akiyama@aist.go.jp}

\author{Mitaro Namiki}
\affiliation{%
  \institution{Tokyo University of Agriculture\\ and Technology}
}
\email{namiki@cc.tuat.ac.jp}

\begin{abstract}
  Applications in the AI and HPC fields require much memory capacity,
  and the amount of energy consumed by main memory of server machines is ever increasing.
  Energy consumption of main memory can be greatly reduced by applying approximate computing in exchange for increased bit error rates.
  AI and HPC applications are to some extent robust to bit errors because small numerical errors are amortized by their iterative nature.
  However, a single occurrence of a NaN due to bit-flips corrupts the whole calculation result.
  The issue is that fixing every bit-flip using ECC incurs too much overhead
  because the bit error rate is much higher than in normal environments.
  We propose a low-overhead method to fix NaNs when approximate computing is applied to main memory.
  The main idea is to reactively repair NaNs while leaving other non-fatal numerical errors as-is to reduce the overhead.
  We implemented a prototype by leveraging floating-point exceptions of x86 CPUs,
  and the preliminary evaluations showed that our method incurs negligible overhead.
\end{abstract}

\maketitle

\section{Background}
Energy saving of data centers and server machines is an important issue.
According to a statistic~\cite{epa2007}, data centers spent 1.5 \% of the total energy consumption of the US in 2006 and 30 \% of the energy consumption was due to the IT equipment (e.g. servers).
As demands for HPC, big data analytics, and AI-related large scale workloads executed in larger scale data centers become stronger, the energy consumption of data centers is expected to increase.
\renewcommand{\thefootnote}{}\footnote{*Work done while Shinsuke Hamada was an intern at National Institute of Advanced Industrial Science and Technology in collaboration with Soramichi Akiyama.}

Large memory usage of HPC and AI-related workloads causes the main memory capacity of server machines to be increased.
This results in the proportion of energy consumed by main memory to the overall energy consumption of a server machine to also increase.
Some report that 25\% -- 40\% of the overall energy of a server machine is consumed by its memory~\cite{Barroso2013, Meisner2009}.
Thus, reducing energy consumption of main memory greatly reduces overall energy consumption of server machines.
This motivates much work such as ones that optimize the conventional DRAM~\cite{Thang2014, Lee2017, Khan2017} and others that utilize non-volatile memory devices~\cite{koshiba2017, emre2013, Sun2017}.

\section{Approximate Memory}
\subsection{Overview}
Approximate computing is a new approach to achieve better efficiency (e.g. reduced energy consumption, increased capacity, wider bandwidth, lower latency) by
lowering computation accuracy and allowing errors mixed into calculated results.
Although the integrity of calculation results and stored data are lost,
approximate computing achieves much efficiency at a level that cannot be realized when accurate computation is guaranteed.
For example, an approximate network achieves high bandwidth and low latency by sacrificing error-free transmission~\cite{fujiki2017},
and an approximate storage increases its capacity without additional flash chips in return of an increased bit-error rate~\cite{Guo2016}.

Applying approximate computing to main memory is a promising way to reduce energy consumed by main memory.
Lowering DRAM refresh rates increases the number of bit-flips, but can greatly reduce the energy consumption.
Applying a lower refresh rate for error-robust DRAM rows saves 16.1\% of energy consumed by memory in an 8-core machine~\cite{Liu2012},
and applying a lower refresh rate for non-critical data saves 20\% -- 25\% of energy consumed by memory in a mobile device~\cite{Liu2011}.
Especially for numerical applications that appear in the HPC and AI fields, the effect of bit-flips can be amortized by their iterative nature.
Fang {\it et al.}~\cite{Fang2017} claim that many HPC applications yield acceptable results even if a few pointers are broken and the pointed data cannot be fetched,
and Liu {\it et al.}~\cite{Liu2011} claim that numerical applications such as a ray shader are robust to bit-flips.

\subsection{\label{section:challenge}Challenge}
\begin{figure}[t]
  \begin{center}
    \includegraphics[width=0.7\columnwidth]{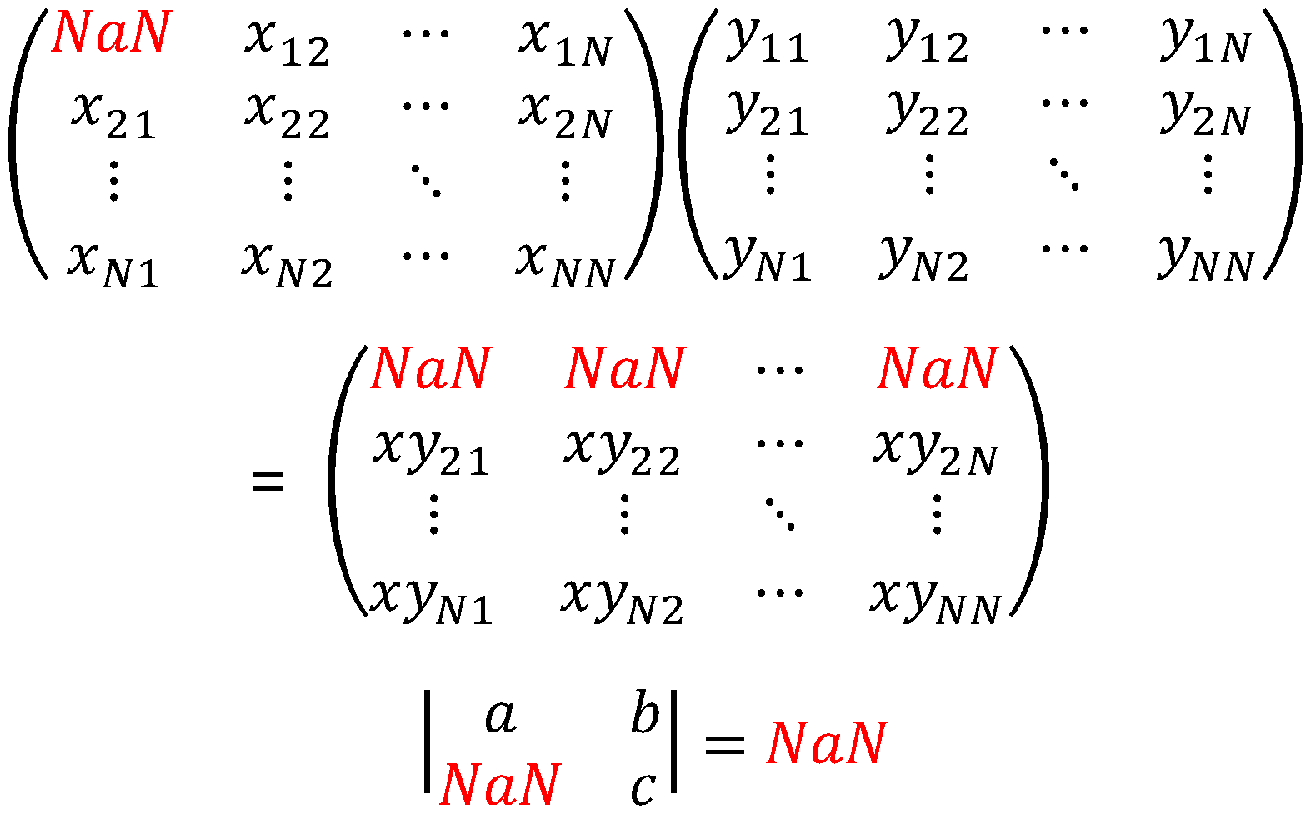}
    \caption{Examples of amplified effect of NaNs. Top: A single NaN in a matrix-matrix multiplication makes all values in a row of the result meaningless.
      Bottom: The determinant of a matrix containing a single NaN is also a NaN.}
    \label{matrix_nan}
  \end{center}
\end{figure}

Although numerical applications are to some extent robust to bit-flips, the challenge is that a single occurrence of a NaN (Not-a-Number) makes calculation results meaningless.
In particular, the effect of a NaN is amplified by matrix operations, which are common across numerical applications.
Figure~\ref{matrix_nan} shows examples where calculation results become meaningless by a single NaN.
In a matrix-matrix multiplication shown in the top of the figure, a single NaN changes all the elements of a row of the result to NaNs.
Another example in the bottom of the figure
shows the determinant of a matrix including a NaN is also a NaN.
These examples show that an occurrence of a NaN due to bit-flips is a fatal problem for numerical applications that use floating-point numbers.
Changing a floating-point number to a NaN requires to flip all bits of the exponent part to 1,
but we believe this happens with a non-negligible probability in a future approximate computing environment with high memory density and high error-rate.
Also, using short bit-widths numbers is encouraged for some AI algorithms~\cite{Zhou2016}, where the number of bits of the exponent part is much smaller than the one of a 64-bit number.

Error-Correcting Code (ECC) memory is a popular method for detecting and correcting bit-flips in main memory.
Although ECC memory is widely adopted in the current server market, it cannot be used to fix NaNs in approximate memory.
ECC memory adds several parity bits to stored data.
A write to ECC memory requires encoding the written data with parity bits, and a read from ECC memory requires decoding the data.
Therefore, enabling the correction of a large number of bits by ECC memory
greatly penalizes memory throughput due to the encoding and decoding overhead~\cite{hirofumi2017}.

To apply approximate computing to main memory for numerical applications,
we need a  {\it flexible and low overhead} approach to prevent calculation results from becoming meaningless upon the occurrence of a NaN.
The requirement for flexibility comes from the fact that a value to which a NaN should be fixed depends on each application.
This means that a hardware mechanism that defines the result of any operation involving a NaN to be a pre-defined value (e.g. 0) does not work.
The requirement for low overhead comes from our final goal, i.e., reducing energy consumption of server machines.
In order to maximize the energy efficiency, the overhead of our method must be as small as possible.
To fulfill these requirements, this paper proposes a software-based flexible method to deal with NaNs with low overhead.

\section{Proposal}
\subsection{Main Idea: Reactive NaN Repair}
The main idea of our method is to {\it reactively} {\it repair} NaNs that appear due to bit-flips.
{\it Repairing} a NaN means to modify the bit-pattern of a NaN and change it to a legal value as a floating-point number.
For repairing NaNs, two types of methods are possible:
{\it proactive} methods that deal with every bit-flips regardless of the actual value (it might be a NaN, but it might also be a drifted floating-point number that can be amortized by the application),
and {\it reactive} methods that deal with bit-flips only after they incur a NaN and a program failure.
An advantage of proactive methods is that they can repair any bit-flips by hardware and software does not even notice that bit-flips have occurred.
A disadvantage is that the performance is greatly reduced because it must check every bit of large memory capacity.
On the other hand, an advantage of reactive methods is that the overhead is small because the repairing procedure is executed only when a program failure occurs.
A disadvantage is that they cannot deal with unexpected situations.
For example, our method cannot repair invalid pointers that appear due to bit-flips.
As described in Section~\ref{section:challenge}, proactive methods such as ECC is not suitable in our scenario due to its large overhead to deal with high error rate and a reactive method to is mandatory.

We propose a reactive method to deal with NaNs by leveraging the fact that
floating-point arithmetic instructions invoke exceptions when an operand is a NaN.
The method is achieved by catching a floating-point exception raised by a CPU and modifying bit patterns of the NaN that caused the exception to a legal floating-point value.
Our method consists of two repairing mechanisms:
\begin{enumerate}
\item Register-repairing mechanism: The direct cause of an exception is a NaN in a register.
  The NaN is repaired to let the application continue, otherwise the application dies due to the exception.
\item Memory-repairing mechanism: Because bit-flips occur inside main memory, a NaN in a register has its origin in memory.
  We repair the NaN in main memory as well to suppress occurrences of further floating-point exceptions to reduce the overhead.
\end{enumerate}

\subsection{Design}
We explain basic implementation design based on the x86 CPU architecture.
Note that the method does not rely on any x86-specific feature, so we believe it can easily be ported to other architectures such as ARM (see Section~\ref{section:discussion} for the details).

\begin{figure}[t]
  \begin{center}
    \includegraphics[width=0.75\columnwidth]{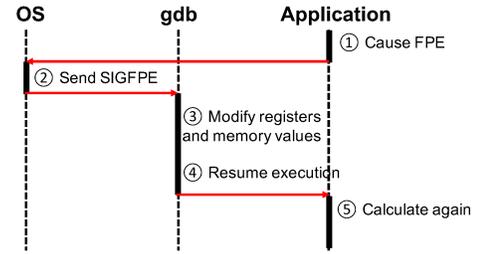}
    \caption{The procedure of our proposed method.
      \ctext{1} The application causes a floating point exception when it touches a NaN.
      \ctext{2} The OS sends a SIGFPE signal to the application.
      \ctext{3} The signal is stolen by our mechanism and the NaN is repaired.
      \ctext{4} The application execution is resumed.
      \ctext{5} The application continues running as if nothing has happened.}
    \label{fig:flow}
  \end{center}
\end{figure}

\begin{figure*}[t]
  \begin{center}
    \includegraphics[width=1.6\columnwidth]{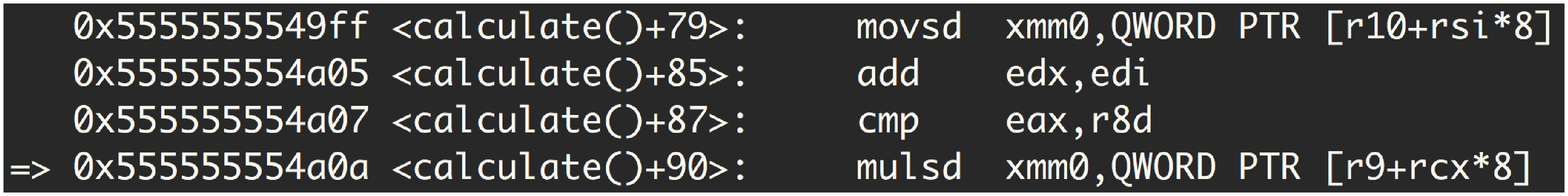}
    \caption{\label{fig:sigfpe}Execution context when a SIGFPE signal is raised at a floating-point instruction (mulsd).}
  \end{center}
\end{figure*}

\begin{figure}[t]
  \begin{center}
    \includegraphics[width=0.65\columnwidth]{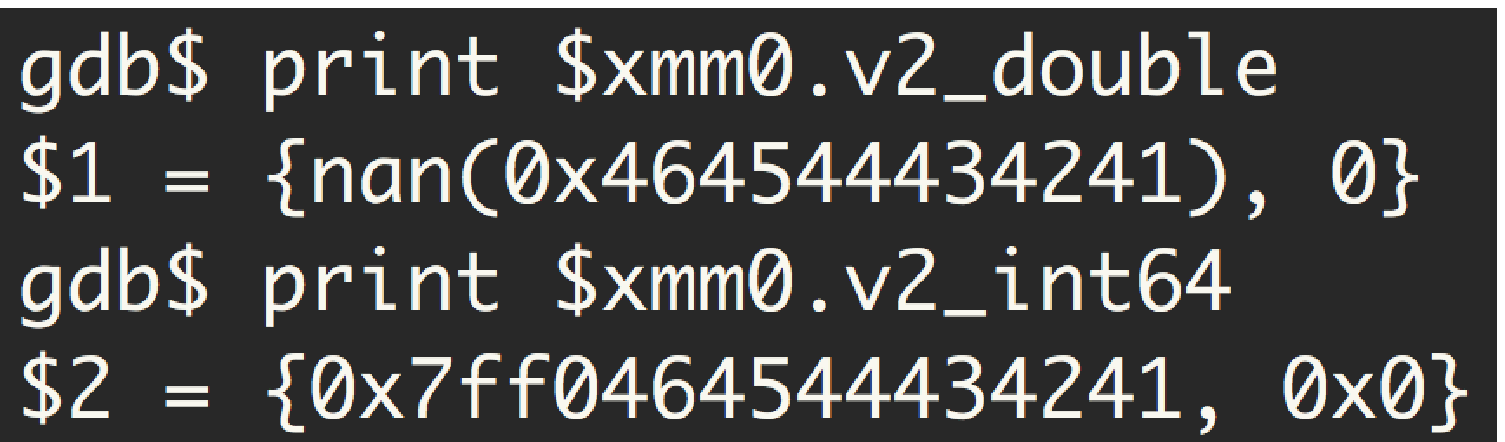}
    \caption{\label{fig:xmm0}Investigating {\bf xmm0} register by gdb}
  \end{center}
\end{figure}

The proposed method is achieved by interrupting program execution when it executes a floating-point instruction whose operand is a NaN
and modify the contents of registers and memory regions that the program uses.
In the x86 architecture, a floating-point arithmetic operation causes an exception when an operand is a NaN.
The exception is handled by an OS to let the software decide what to do.
Linux delivers a SIGFPE signal to the program that has executed the floating-point operation.
We modified this behavior so that the register and memory contents are repaired before the program restarts its execution.

For our prototype implementation, we use \verb|gdb| to ``steal'' signals from the workload process.
We also use \verb|gdb python| to control \verb|gdb| by pre-written scripts.
Please carefully note that this choice is not mandatory but for simplicity,
and one can choose more general mechanisms such as the \verb|ptrace| system call or modifying signal handlers of the OS.
A target workload is attached with \verb|gdb| and executed.
Figure~\ref{fig:flow} shows the procedure of the proposed method.
When the target workload executes a floating-point operation with a NaN in its operand caused by bit-flips,
a floating-point exception is incurred by the CPU and a SIGFPE signal is sent to the workload from the OS (\ctext{1}).
Because \verb|gdb| is attached to the workload, it can ``steal'' the SIGFPE signal from the workload (\ctext{2}).
Then, \verb|gdb| repairs NaNs in registers and main memory by using the two methods described in the following sections (\ctext{3}).
Finally, \verb|gdb| resumes the workload and the workload can continue the execution as if nothing has happened (\ctext{4} and \ctext{5}).

The overhead caused by attaching \verb|gdb| is incurred only when the signal handling and NaN repairing procedures are executed.
Thus, except when floating-point exceptions are being handled, no overhead is incurred to the target workload.
Furthermore, the number of times floating-point exceptions are incurred by a single NaN is limited to only 1 by our method to repair memory contents (Section~\ref{section:memory_repair}), resutling in negligile overhead of our proposed method.

\subsection{Register-Repairing Mechanism}
This section explains how to identify a NaN in a register and how to repair it.
When \verb|gdb| catches a SIGFPE signal, the instruction pointer of the CPU points to a floating-point instruction that has a NaN as its operand.
Figure~\ref{fig:sigfpe} shows the context of a workload when a SIGFPE signal is caught.
The {\bf mulsd} instruction at the 4$^{\rm th}$ line (which is located at the 90$^{\rm th}$ byte of the \verb|calculate| function) is pointed to by the instruction pointer,
thus it is clear that the {\bf xmm0} register contains a NaN caused by bit-flips.
Figure~\ref{fig:xmm0} shows the value of {\bf xmm0}, which is interpreted as a NaN.
The workload restarts its  execution after \verb|gdb| has repaired the NaN in the floating-point register to a legal value as a floating-point number.
The choice of an actual value to which a NaN is fixed is a next question, but we leave it as future work (see Section~\ref{section:values_to_be_fixed} for the details).
The workload can resume the execution without knowing that the floating-point exception has occurred because it is transparently handled by \verb|gdb| and not passed to the workload process.

\begin{figure}[t]
  \begin{center}
    \includegraphics[width=0.65\columnwidth]{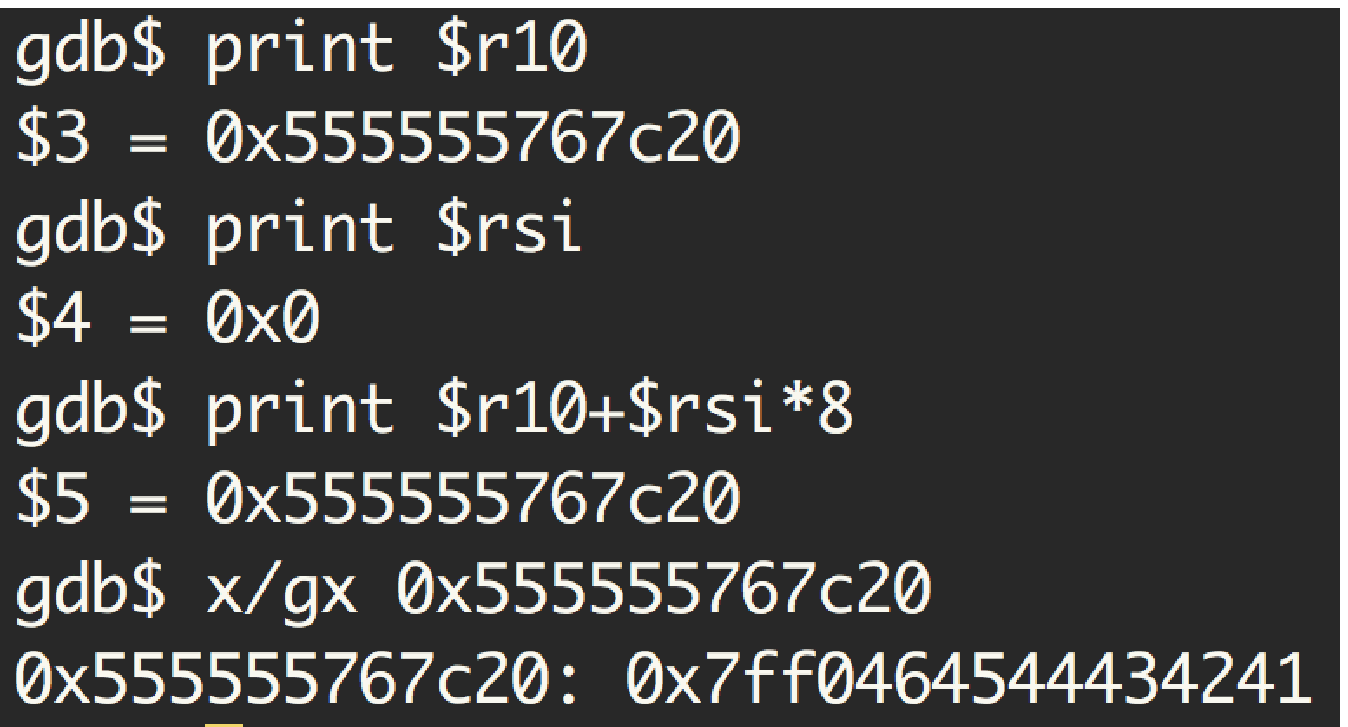}
    \caption{\label{fig:memory}Finding the memory address of a NaN using the operands of the movsd instruction in Figure 3.}
  \end{center}
\end{figure}

\subsection{\label{section:memory_repair}Memory-Repairing Mechanism}
This section describes how to repair NaNs that exist in main memory.
As we have described, a workload can continue as if nothing has happened after NaNs that exist in registers are repaired.
However, in an environment where main memory is approximated, a NaN in a register has its origin in main memory.
Even after the NaN is repaired on the register, if it is loaded from main memory to a register again, a floating-point exception and the overhead to repair it are also incurred again.
To solve this issue, the proposed method repairs the NaN in memory as well.

To repair a NaN in main memory, the memory address where the NaN is stored is required.
When a workload triggers a SIGFPE, the execution context contains a floating-point instruction whose operand is a NaN.
We identify the memory address of the NaN by analyzing the workload binary.
The {\bf mov} instruction that loads the NaN from main memory is searched for and the address is identified by investigating the operands of the {\bf mov} instruction.
Suppose a SIGFPE is raised when the {\bf mulsd} instruction at the 4$^{\rm th}$ line in Figure~\ref{fig:sigfpe} is executed, and a NaN is stored in {\bf xmm0}.
By disassembling the binary, we can find that the content of {\bf xmm0} is loaded from main memory with the {\bf movsd} instruction in the 1$^{\rm st}$ line.
It loads a QWORD value from the memory address of {\bf r10} $+$ {\bf rsi} $\times$ 8.
Figure~\ref{fig:memory} shows that the value of {\bf r10} $+$ {\bf rsi} $\times$ 8 is \verb|0x555555767c20|
and the value stored in this address is \verb|0x7ff0464544434241|, which is a NaN when interpreted as a 64-bit floating-point number.

The execution context at the time a SIGFPE is raised includes the values of registers of that time only.
This raises an issue that the memory address of a NaN is not necessarily identifiable due to two reasons:
(1) The corresponding {\bf mov} might not be found because a binary is not always back-traceable.
For example, a conditional branch cannot be back-traced by using only the program binary.
(2) The operands of the corresponding {\bf mov} might not be preserved.
For example in Figure~\ref{fig:sigfpe}, if either of {\bf r10} or {\bf rsi} is modified somewhere between the {\bf mulsd} and the {\bf mov}, it is not possible to identify the address from which the NaN is read.
Note that recording detailed traces using dynamic binary analysis tools such as Intel PIN~\cite{Intel_PIN} can solve this issue, but the overhead is too heavy.

\begin{figure}[t]
  \begin{center}
    \includegraphics[width=0.85\columnwidth]{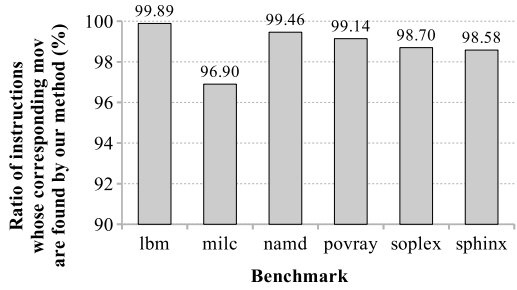}
    \caption{Ratio of floating-point arithmetic instructions whose corresponding mov's are found from the execution context and the binary.}
    \label{figure:graph}
  \end{center}
\end{figure}

\begin{table}[t]
  \caption{List of floating-point instructions (arithmetic, mov-related) covered by our memory-repairing mechamism.}
  \label{table:sseinst}
  \centering
  \begin{tabular}{|c|c|}
    \hline
    arithmetic & addss, addpd, addsd, addps \\
    & mulss, mulpd, mulsd, mulps, \\
    & subss, subpd, subsd, subps, \\ 
    & divss, divpd, divsd, divps \\ \hline
    mov & movss, movsd, movd \\ \hline
  \end{tabular}
\end{table}

We confirmed that the two cases above are rare enough so that we can repair NaNs in memory with a probability exceeding 95\%.
This does not mean that the other 5\% of NaNs immediately crash a workload,
but rather they are fixed with the register-repairing mechanism with higher overhead.
We investigated the binaries of SPEC CPU 2006 benchmarks that mostly calculate floating-point arithmetic.
They are built in the environment shown in Table~\ref{table:eval} with the \verb|-O2| compiler option.
We disassembled the binaries and searched for instructions related to floating-point arithmetic such as {\bf addss}.
For each instruction $I$, we searched for the corresponding {\bf mov}-related instruction $M$ that loads the operands of $I$ from main memory.
Table~\ref{table:sseinst} lists the floating-point arithmetic instructions and {\bf mov}-related instructions that we cover. 
The corresponding {\bf mov}-related instruction $M$ of a floating-point arithmetic instruction $I$ is referred to as {\it found}
if $M$ exists in a back-traceable place from $I$ (meaning that $I$ and $M$ are in the same function and there is no conditional branch in-between),
and registers used as $M$'s operands are not modified between $I$ and $M$.
Figure~\ref{figure:graph} shows the percentage of floating-point arithmetic instructions whose corresponding {\bf mov} is found,
calculated against all floating-point arithmetic instructions in the binary.
The results show that the corresponding {\bf mov} is found for more than 95\% of the floating-point arithmetic instructions we deal with.
It means that a NaN in main memory can be repaired in most cases and that the overhead incurred by the repairing procedure can be successfully suppressed by our memory-repairing mechanism.

\section{Preliminary Evaluation}
\begin{table}[t]
  \centering
  \caption{\label{table:eval}Evaluation environment}
  \begin{tabular}{|c|c|}
    \hline
    OS &  Ubuntu 16.04 (Linux 4.4.0-92-generic) \\
    CPU & Intel Core i7 870 (2.93GHz) \\
    gcc & 5.4.0 \\
    gdb & 7.11.1 \\
    python & 2.7.12 \\
    \hline
  \end{tabular}
\end{table}

As a preliminary evaluation, we measured the overhead of our method.
We apply the proposed method to an application that executes a matrix-matrix multiplication and measured the execution time,
because matrix-matrix multiplication is one of the most common components of various numerical applications.
The application allocates two matrices of size $N \times N$ and multiplies them.
We compare execution time of the application in three conditions:
(1) it is executed normally (2) a NaN is intentionally injected and it is repaired by our register-repairing mechanism,
and (3) a NaN is intentionally injected and it is repaired both by our register- and memory-repairing mechanisms.
A NaN is injected in to one of the two matrices after their initialization to mimic an occurring of a NaN by bit-flips.
We measured the execution time 10 times and took the average.
Table~\ref{table:eval} shows the evaluation environment.

\begin{figure}[t]
  \centering
  \includegraphics[width=0.8\columnwidth]{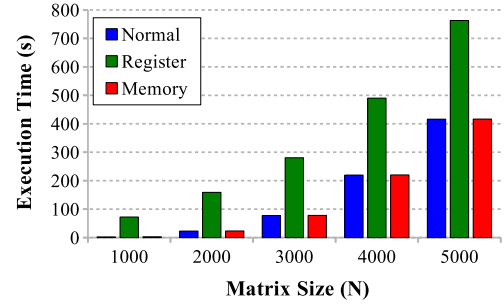}
  \caption{Elapsed time of matrix-matrix multiplication.
    Normal: executed normally (no NaNs).
    Register: a NaN is injected and repaired by the register-repairing mechanism.
    Memory: a NaN is injected and repaired both by the register- and memory-repairing mechanisms.
  }
  \label{figure:pre_eval}
\end{figure}

Figure~\ref{figure:pre_eval} shows the evaluation results.
The $x$ axis shows the size of the matrices ($N$) and the $y$ axis shows the execution time.
The label {\bf normal} refers the results when the application is executed normally,
{\bf register} refers the results when a NaN is injected and repaired by our register-repairing mechanism,
and {\bf memory} refers the results when a NaN is injected and repaired both by our register- and memory-repairing mechanisms.
The results show that by combining register- and memory-repairing mechanisms,
the overhead to repair a NaN can be suppressed to a negligible amount.
We confirmed the same trend for a matrix-vector multiplication application as well,
but the detailed numbers are omitted due to the space limit.

Table~\ref{table:sigfpe} shows the number of SIGFPEs occurred in each method.
When only the register-repairing mechanism is applied (``register'' in the table), a single NaN in memory is loaded to a register again and again
and the overhead by the repairing procedure is incurred every time a SIGFPE is raised.
On the other hand, when both the register- and memory-repairing mechanisms are applied (``memory'' in the table),
a NaN in memory is repaired at the first occurrence of a SIGFPE, resulting a negligible amount of overhead.

\section{Discussions}

\begin{table}[t]
  \caption{Number of SIGFPEs incurred in each method for matrix-matrix multiplication.}
  \label{table:sigfpe}
  \begin{center}
    \begin{tabular}{|c|c|c|c|c|c|} \hline
      Matrix Size  & 1000 & 2000 & 3000 & 4000 & 5000 \\ \hline
      Register     & 1000 & 2000 & 3000 & 4000 & 5000 \\
      Memory       & 1    & 1    & 1    & 1    & 1    \\ \hline
    \end{tabular}
  \end{center}
\end{table}

\subsection{\label{section:discussion}Applicability to Other Architectures}
ARM CPUs have been gathering attention especially by HPC users thanks to their low-energy characteristics~\cite{Rajovic2014}.
We have shown a prototype implementation of our method on x86 CPUs,
but we believe porting it to ARM is straight-forward because it relies on no x86 specific mechanism.
An ARM CPU raises an exception when it applies an arithmetic operation to a NaN,
which switches the execution path to an exception handler of the OS.
After the exception handler for SIGFPE is invoked, the same procedure of our current implementation can be applied to ARM CPUs as well.

GPUs are the key component both in the HPC and AI fields for acceleration.
Applying our method to GPUs requires two features: general purpose exception handling in GPUs and coherent shared memory across a CPU and a GPU.
The latter is supported in recent NVIDIA GPUs so that data (e.g. a NaN) in GPU memory can be easily modified from a CPU.
General purpose exception handling is a big challenge
because a GPU has 100X to 1000X larger number of hardware threads than a CPU and saving all states upon an exception is infeasible.
Researchers tackle this issue by reducing the number of states required to be saved for an exception handling~\cite{Menon2012}.
We anticipate that these evolutions of recent GPUs will help in applying our method to GPUs as well in the future.

\subsection{\label{section:values_to_be_fixed}Values to which NaNs are fixed}
We leave how to determine an actual value to which a NaN is fixed as future work because of two reasons:
(1) it is orthogonal to how to fix the NaN with low overhead, which is the focus of this paper, and
(2) how to determine the value is a hot research topic itself.

LetGo~\cite{Fang2017} mitigates SIGSEGVs caused by bit-flipped pointers by modifying the program state so that the program can continue as if it had read a 0 from the pointer.
They claim that a 0 makes many HPC applications converge to reasonable results.
However, some applications have divisions, in which case using 0s causes another failure.
For instance, a typical LU decomposition algorithm contains divisions.
{\it Pivot selection} is a common way to avoid divisions by 0s for LU decomposition,
but it cannot deal with a case where a value can be changed to a 0 anytime.
For example, a bit-flip may occur right after a value has been confirmed to be non-0.

Li {\it et al.}~\cite{Li2017} analyze how soft errors affect the accuracy of deep learning.
They report that (1) a bit-flip on the sign bit does not affect the accuracy because the values are distributed around 0 with a symmetric distribution in typical deep neural networks, and (2) the effect of one bit-flip is larger when a wider bit-width is used to represent a floating-pointer number.
Although this knowledge is specific to deep learning, it implies that the choice of the value to which a NaN is fixed requires much consideration depending on each target workload.

\section{Related Work}
To the best of our knowledge, this paper is the first to handle NaNs that appear when approximate computing is applied to main memory.
We review existing work that have tackled similar problems in the context of fault tolerance.

LetGo~\cite{Fang2017} deals with bit-flips in pointer variables under a condition of high bit error rate due to high memory density of data centers.
When a program dereferences a bit-flipped pointer and is about to crash, LetGo catches the SIGSEGV (or SIGBUS) signal.
It modifies the program context and registers to let the program go as if the value pointed by the dereferenced pointer was a 0.
In comparison, our method is more advanced because of our memory repairing mechanism
and the quantitative analysis that proves its feasibility.

Bosilca {\it et al.}~\cite{Bosilca2009} propose an approach called {\it Algorithm-Based Fault Tolerance (ABFT)} that targets soft errors in HPC applications.
The key idea is to embed parities into calculated data itself so that soft errors can be detected by software and the calculation can be retried when an error is detected.
Although ABFT is very practical in the sense that it does not rely on any hardware feature,
retrying whole calculation is not suitable for our purpose because it greatly reduces energy efficiency.

Carbin {\it et al.}~\cite{Carbin2011} propose Jolt, which detects when a program enters an infinite loop and allows it to escape from the loop. 
Jolt is attached to a program to monitor its progress by recording the program states at the start of each loop iteration.
If the recorded states in two consecutive loop iterations are the same, it means that the program is trapped by an infinite loop.
A trap by an infinite loop is another possible failure of programs by approximate computing besides occurrences of NaNs.
Because the instrumentation overhead of Jolt is small, it is a good candidate for mitigating infinite loops caused by bit-flips in approximate computing.

\section{Conclusion}
Energy consumption of main memory in server machines are increasing greatly as larger and larger memory capacity is required by large scale AI and HPC workloads.
Applying approximate computing to main memory is a promising way to reduce energy consumption of main memory,
and AI and HPC applications are expected to be robust to the increased bit error rates.
However, a single occurrence of a NaN due to bit-flips corrupts the whole calculation result,
and fixing every bit-flip using ECC incurs too much overhead.
We proposed a reactive method that repairs only NaNs while leaving other non-fatal numeric errors as-is to reduce the overhead.
Our preliminary evaluation showed that the overhead by our method is negligible.

\begin{acks}
  We thank the anonymous reviewers for their constructive suggestions to improve this work.
  We also thank Dr. Jason Haga (AIST) for giving fruitful comments.
\end{acks}

\bibliographystyle{ACM-Reference-Format}
\bibliography{sfma2018}

\end{document}